\documentclass[journal,onecolumn]{IEEEtran}
\usepackage{setspace}
\usepackage{xcolor,soul,framed} 

\colorlet{shadecolor}{yellow}
\usepackage[pdftex]{graphicx}
\graphicspath{{../pdf/}{../jpeg/}}
\DeclareGraphicsExtensions{.pdf,.jpeg,.png}

\usepackage[cmex10]{amsmath}
\usepackage{array}
\usepackage{mdwmath}
\usepackage{mdwtab}
\usepackage{booktabs} 
\usepackage{eqparbox}
\usepackage{url}
\usepackage{rotating}
 \usepackage[caption=false,font=footnotesize]{subfig}
\usepackage{algpseudocode}
\usepackage{url}
\usepackage{amsmath, amssymb}
\usepackage{algorithm}
\usepackage{cite}
\usepackage{graphicx}
\begin{document}

\bstctlcite{IEEEexample:BSTcontrol}
    \title{Pinching Antenna-Aided Spatial Multiplexing: Transceiver Design and Performance Analysis}
  \author{Ruijie Li, 
       Yue Xiao, Shuaixin Yang,
      Gang Wu, Xianfu Lei
      and Ming Xiao

  \thanks{R. Li, Y. Xiao, S. Yang and G. Wu are with the National Key Laboratory of Wireless Communications, University of Electronic Science and Technology of China (UESTC), Chengdu 611731, China (e-mail: 
\mbox{ruijieli@std.uestc.edu.cn}, \mbox{xiaoyue@uestc.edu.cn},
\mbox{shuaixin.yang@foxmail.com}, \mbox{wugang99@uestc.edu.cn}).}
  \thanks{X. Lei is with the School of Information Science and Technology, Southwest Jiaotong University, Chengdu 610031, China (e-mail:
\mbox{xflei@swjtu.edu.cn}).}%
   \thanks{M. Xiao is with the Department of Information Science and Engineering,
Royal Institute of Technology (KTH), 100 44 Stockholm, Sweden (e-mail:
\mbox{mingx@kth.se}).}
}


\maketitle

\begin{abstract}
In this paper, a novel pinching antenna-aided spatial multiplexing (PASM) architecture is conceived, which intrinsically amalgamates the benefits of flexible radiating element placement with radio-frequency (RF) chain transmission. Specifically, we leverage the deterministic phase variation along dielectric waveguides as a zero-power phase-control mechanism, where each waveguide fed by a single RF chain drives multiple pinching antennas (PAs) acquiring position-dependent phase shifts. Then, the PASM propagation environment is characterized by a realistic channel model encompassing Rician small-scale fading, correlated shadowing, and large-scale path loss. Based on this, a low-complexity vector approximate message passing (VAMP) detector is conceived, which exploits a waveguide-structured prior for jointly processing the signals associated with all PAs. Moreover, we derive an analytical upper bound on the bit error rate (BER) for the maximum likelihood (ML) detector to quantify the achievable performance limits. Finally, our simulation results demonstrate that the proposed PASM architecture achieves substantial signal-to-noise ratio (SNR) gain over the conventional phase-shifter-aided spatial multiplexing (PSSM), while the VAMP detector strikes an attractive trade-off between the system performance and computational complexity.
\end{abstract}


%
\IEEEpeerreviewmaketitle


\section{Introduction}

Sixth-generation (6G) wireless networks~\cite{wang2023road,10745905} are envisioned to support ultra-high data rates, ultra-reliable low-latency connectivity, and massive device access under stringent cost, energy, and hardware constraints. In this context, multi-antenna technologies have been recognized as a pivotal enabler for improving both the spectral and energy efficiency in 6G~\cite{bjornson2024enabling,9136592}. By harnessing the available spatial degrees of freedom, these techniques offer substantial beamforming, spatial multiplexing, and interference mitigation gains, which constitute the foundation of modern cellular standards~\cite{heath2016overview,bjornson2017massive}. Unfortunately, as the antenna arrays scale from massive to gigantic~\cite{heath2016overview}, the conventional architecture relying on a single radio-frequency (RF) chain per antenna incurs prohibitive hardware cost, power consumption, and signal processing complexity, especially in the millimeter-wave (mmWave) and terahertz (THz) bands~\cite{rappaport2019wireless,zhang2019prospective,venkatesan2017mmwaveMIMO}. Against this background, substantial research attention has been dedicated to low-RF-chain multi-antenna architectures~\cite{molisch2017hybrid} capable of harvesting significant spatial gains, albeit with significantly reduced hardware overhead.

To circumvent the RF chain bottleneck, a diverse suite of reduction techniques has been advocated. Specifically, hybrid analog-digital beamforming~\cite{el2014spatially,sohrabi2016hybrid} invokes a limited number of RF chains combined with phase shifter networks for approximating fully digital precoding, thereby striking an attractive trade-off between the attainable performance and hardware cost. Furthermore, spatial modulation (SM)~\cite{mesleh2008spatial,di2011spatial} and its generalized variants~\cite{yang2014design,di2013spatial} convey implicit information via the antenna indices by activating only a few RF chains, hence enhancing the spectral efficiency without linearly scaling the RF hardware. Additionally, antenna selection (AS)~\cite{sanayei2004antenna} dynamically activates optimal antenna subsets for maximizing the effective signal-to-noise ratio (SNR) under a stringent RF chain budget. However, although those above-mentioned approaches achieve substantial RF chain savings, they often suffer from multiplexed gain erosion or require complex analog beamforming networks relying on numerous active phase shifters.

Recently, phase-shifter-aided spatial multiplexing (PSSM) has been developed as a promising low-RF-chain candidate~\cite{chen2023phase}. By connecting a single RF chain to multiple antennas via agile phase shifters, PSSM can maintain the high spectral efficiency without relying on antenna index modulation. It has been demonstrated in~\cite{chen2023phase} that PSSM outperforms SM and AS in the high spectral-efficiency regime, whilst requiring significantly fewer RF chains than the vertical Bell Labs layered space-time (V-BLAST) architecture~\cite{hassibi2002high}. Nevertheless, conventional PSSM implementations rely on discrete active phase shifters, which suffer from insertion loss and quantization errors~\cite{payami2018hybrid,tarboush2021teramimo}. Consequently, as the antenna arrays scale towards the gigantic regime envisioned for 6G, the cost of employing numerous active phase shifters becomes prohibitive~\cite{hoydis2013massive,xue2020energy}, hence limiting the scalability of PSSM systems.

Consequently, in the quest to liberate multi-antenna systems from these hardware constraints, the research focus has progressively shifted from conventional fixed-position arrays towards flexible antenna architectures. These emerging technologies have been advocated to proactively reconfigure wireless channels rather than merely adapting to them passively~\cite{gong2020toward}.
Specifically, reconfigurable intelligent surfaces (RISs) \cite{basar2019wireless,wu2019towards,pan2022overview} employ large numbers of nearly passive reflecting elements whose phase shifts can be electronically adjusted to manipulate the wireless propagation environment, enabling intelligent reflection, highly directional beamforming, interference management, and coverage enhancement in a cost- and energy-efficient manner. Furthermore, fluid antennas (FAs) \cite{wong2020fluid} and movable antennas (MAs) \cite{zhu2023modeling,chen2024movable} physically reposition radiating elements within a small region to harness spatial small-scale fading diversity, hence improving link reliability, providing an additional degree of freedom for load balancing, and offering robustness to user mobility and blockage \cite{ding2025flexible,yang2025flexible}. Notwithstanding their merits, the practical deployment of these existing paradigms is subject to specific physical constraints. To tackle these design challenges, the concept of pinching-antenna systems (PASS) has been introduced as a particularly novel and cost-effective flexible-antenna architecture~\cite{ding2025pinching}. PASS employs a fundamentally different mechanism, where electromagnetic waves are guided by low-loss dielectric waveguides that can be deployed over large areas, and radiation into free space is triggered by small dielectric particles, termed pinching antennas (PAs), which are mechanically attached to the waveguides. PASS is different from other flexible architectures in three important ways. First, by leveraging the mechanical mobility of PAs along the extended waveguides, the radiation points can be physically maneuvered to positions near the users. This capability enables the system to establish strong line-of-sight (LoS) links, thereby effectively mitigating severe large-scale fading~\cite{ding2025flexible,liu2025pinching}. Second, PAs can be added, removed, or repositioned at will, enabling highly scalable reconfigurability in both the number and spatial distribution of radiating elements~\cite{yang2025pinching}. Third, and most significantly for signal processing, the guided waves propagating through the dielectric waveguide naturally experience a deterministic and continuous phase progression. This intrinsic physical phenomenon effectively functions as a natural, zero-power analog phase shifter, dispensing with the cumbersome active phase shifter hardware required by conventional fixed-position arrays~\cite{mu2025pinching}. These unique features have motivated a burgeoning body of research on PASS, ranging from physics-based modeling to beamforming optimization~\cite{wang2025modeling,zhu2025pinching,wang2025pinching,wang2025antenna,ding2025pinching,yang2025paim}.

Despite these advances, most existing PASS-related works characterize PAs primarily as discrete radiation points that extend coverage or provide additional spatial degrees of freedom, while the transceiver design largely adheres to conventional MIMO paradigms. Although the recent pinching antenna-assisted index modulation (PA-IM) scheme~\cite{yang2025paim} validates the merit of PAs as information-bearing resources by embedding bits into the activation indices, its transmission strategy is fundamentally constrained by the limited number of available antenna combinations, thereby prioritizing spatial diversity over multiplexing gain. From a system design perspective, this feature represents a significant underutilized potential: if the deterministic phase evolution along the waveguide could be harnessed in a manner analogous to active phase shifters in PSSM, one could simultaneously achieve RF chain reduction and favorable large-scale channel reconfiguration within a unified architecture.

Motivated by this gap, we conceive a novel PA-aided spatial multiplexing (PASM) architecture. In this framework, each dielectric waveguide functions as a “virtual antenna” fed by a single RF chain, while multiple PAs residing on the waveguide share the common signal but acquire unique, position-dependent phase shifts induced by the guided-wave propagation. This mechanism effectively constructs a composite constellation vector, enabling multi-stream transmission that mimics the functionality of PSSM but dispenses with active phase shifters. Crucially, beyond mere hardware simplification, PASM distinguishes itself by unlocking additional spatial degrees of freedom. The continuous mobility of PAs facilitates a highly dynamic and generalized position index construction, where the specific locations of PAs can be flexibly optimized to not only combat path loss but also actively shape the effective channel and signal constellation. Consequently, this versatile architecture achieves significantly higher spectral efficiency compared to its fixed-position counterparts, striking a compelling balance between hardware complexity and system throughput.
\begin{figure}[t!]
\centerline{\includegraphics[width=1.05\linewidth]{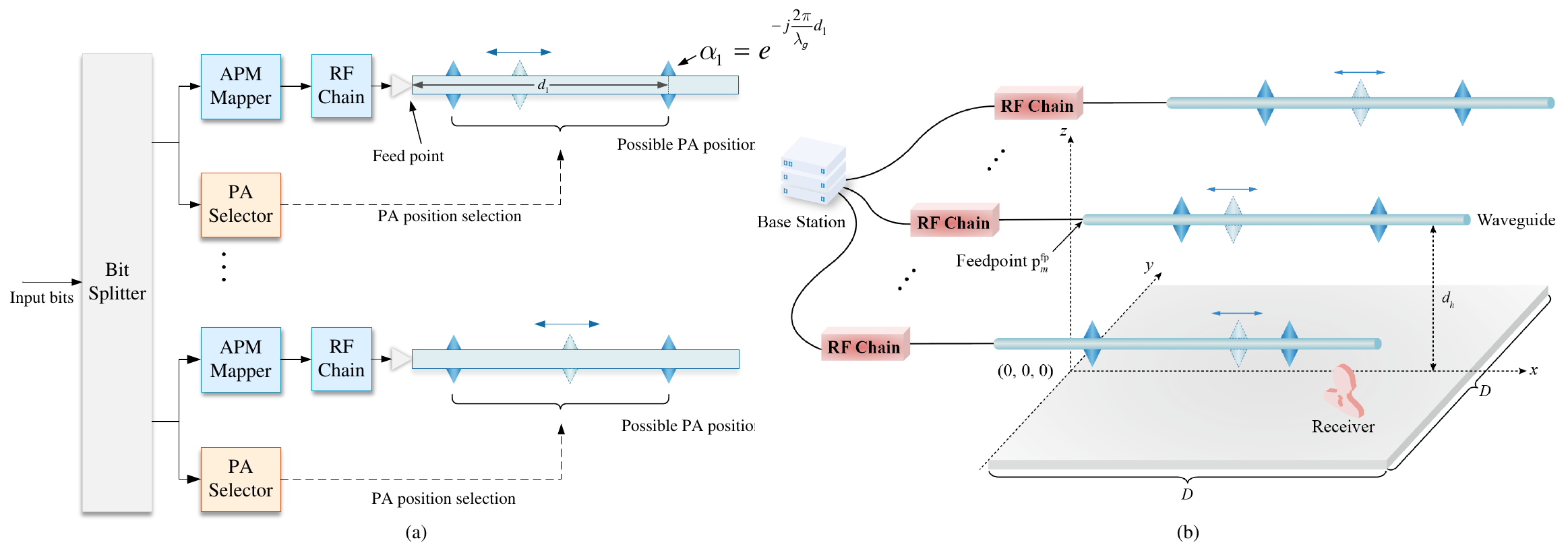}}
\caption{System model of the proposed PASM system: (a) Transmitter architecture and signal flow, where $\alpha_{1}$ represents the waveguide-induced phase shift acquired by the active PA on the waveguide at distance $d_{1}$; (b) Illustration of the deployment and propagation scenario.}
\label{system}
\end{figure}
The main contributions of this paper are summarized as follows:
\begin{itemize}
  \item A novel PASM architecture is developed in which the dielectric waveguide functions as an intrinsic phase-shifter-free multiplexing medium. In this design, a single RF chain feeds multiple PAs that share a common signal but acquire unique position-dependent phases, thereby enabling high-rate spatial multiplexing while simultaneously exploiting the flexible deployment of PASS to mitigate severe large-scale fading.
  \item Building upon the fundamental propagation principles of dielectric waveguides, we establish a system-specific signal and channel model for PASM that captures the waveguide-induced phase progression, Rician small-scale fading, shadow fading, as well as large-scale path loss.
  \item We design a waveguide-structured prior for the PASM transmit vector and intrinsically amalgamate it with a vector approximate message passing (VAMP) detection framework. This bespoke detector strikes a compelling trade-off between the attainable system performance and the computational complexity.
  \item An analytical upper bound on the bit error rate (BER) of the proposed PASM system employing maximum likelihood (ML) detection is derived, the tightness of which is quantified via numerical simulations.
  \item Extensive simulation results demonstrate that the proposed PASM scheme outperforms the conventional PSSM system, particularly in challenging propagation scenarios characterized by severe large-scale fading and spatially correlated channels.
\end{itemize}

The remainder of this paper is organized as follows. Section~II details the system model of the proposed PASM architecture along with the associated channel model. Section~III formulates the factor-graph formulation and conceives the proposed waveguide-level VAMP detector. Section~IV is dedicated to the BER performance analysis of the PASM system, where practical PA deployment patterns as well as waveguide and PA selection strategies are also elaborated on. Numerical results that validate the effectiveness of the proposed scheme and compare it with several benchmark detectors are provided in Section~V. Finally, Section~VI concludes the paper and outlines potential directions for future research.

\textit{Notation:} Scalars, vectors, and matrices are denoted by lower-case, bold lower-case letters, and bold upper-case letters, respectively. $\mathbb{R}$ and $\mathbb{C}$ denote the set of real and complex numbers. The transpose, Hermitian transpose, conjugate, determinant, Euclidean norm, Frobenius norm, trace and vectorization operators are denoted by $(\cdot)^{T}$, $(\cdot)^{H}$, $(\cdot)^{*}$, $|\cdot|$, $\|\cdot\|_{2}$, $\|\cdot\|_{F}$, $\mathrm{Tr}(\cdot)$ and $\mathrm{vec}(\cdot)$, respectively. $\odot$ and $\otimes$ are the Hadamard and Kronecker products. $\oslash$ denotes element-wise division between vectors. diag($\cdot$) and blkdiag($\cdot$) create a diagonal and a block-diagonal matrix from their arguments. An identity matrix of dimension $N\times N$ is denoted by $\mathbf{I}_N$. Furthermore, $\mathcal{CN}(x;\mu,\Sigma)$ denotes a complex Gaussian distribution with mean $\mu$ and variance $\Sigma$. $\mathbb{E}(\cdot)$ and $\mathrm{Var}(\cdot)$ denote the mean and variance, respectively. 
\section{System Model}

The proposed PASM system operates within a square communication region of side length $D$ meters, as illustrated in Fig. \ref{system}(b). The base station is equipped with $N_{\mathrm{wg}}$ spatially separated dielectric waveguides, where each waveguide is fed by an independent RF chain and hosts $N_a$ PAs. During each transmission interval, $N_a$ specific PA positions are dynamically activated on each waveguide, resulting in a total of $N_t^{\mathrm{ant}}=N_\mathrm{wg}\times N_a$ active radiating elements across the entire system. The receiver employs a conventional fixed-position uniform linear array (ULA) comprising $N_r$ elements. A three-dimensional Cartesian coordinate system $(x,y,z)$ is established in Fig. \ref{system}(b) with the waveguides positioned parallel to the $x$-axis at a height of $d_h$ meters above the communication plane.

The fundamental principle of the PASM architecture lies in the joint exploitation of baseband symbol modulation and antenna-side spatial phase modulation to enhance spatial multiplexing gain while reducing hardware complexity. Specifically, for each waveguide, the incoming information bit stream is partitioned into two distinct components: the first subset is mapped to amplitude–and–phase–modulation (APM) symbols, and the other is conveyed by selecting PA activation positions to modulate the transmit antenna phases. The detailed information-bearing strategy will be elaborated in Section II-B, while the selection criteria for PA activation patterns will be discussed in Section IV.

\subsection {Channel Model}

In much of the existing literature on PAs and related spatially reconfigurable architectures, the channel is modeled via the classical Friis transmission equation, which is widely used to describe free-space LoS propagation. Although such models offer considerable analytical convenience by establishing a deterministic relation between antenna displacement and large-scale path loss, they inherently overlook the random nature of practical wireless channels. In particular, multipath propagation and shadow fading, which are ubiquitous in real deployments, especially in dense urban scenarios, are typically ignored.

Moreover, the underlying assumption of a dominant LoS component is often violated in realistic environments, where radio propagation is strongly affected by random blockages and obstructions caused by buildings, vehicles, and terrain. As a result, the significant impact of shadow fading, i.e., the random attenuation induced by large obstacles, is usually not captured, leading to a noticeable mismatch between analytical predictions and measured system performance. It is therefore essential to explicitly model the shadow fading component to obtain credible evaluations of coverage and reliability in complex propagation environments.

To accurately describe these effects, we consider a receiver equipped with a fixed ULA. The spatial coordinates of the $N_r$ receive antennas are collected in the set $\{\mathbf{p}_n^{\mathrm{r}} \in \mathbb{R}^3 \mid n=1,2,\ldots,N_r\}$, where $\mathbf{p}_n^{\mathrm{r}} = [x_n^{\mathrm{r}},\,y_n^{\mathrm{r}},\,z_n^{\mathrm{r}}]^T$. On the transmit side, the base station employs $N_{\mathrm{wg}}$ waveguides with multiple PAs deployed on each of them. The position of the $i$-th activated PA on the $m$-th waveguide is denoted by $\mathbf{p}_{m,i}^{\mathrm{tx}} = [x_{m,i}^{\mathrm{tx}},\,y_{m,i}^{\mathrm{tx}},\,z_{m,i}^{\mathrm{tx}}]^T$, and the injection point of the RF chain (i.e., the feed point of the $m$-th waveguide) is given by $\mathbf{p}_m^{\mathrm{fp}} = [x_m^{\mathrm{fp}},\,y_m^{\mathrm{fp}},\,z_m^{\mathrm{fp}}]^T$.

Within this framework, the channel between the $m$-th waveguide and the receive array is represented by the matrix $\mathbf{H}_m$. The $(j,i)$-th entry of $\mathbf{H}_m$, denoted by $h_{j,i}^{(m)}$, corresponds to the channel coefficient between the $i$-th activated PA on the $m$-th waveguide and the $j$-th receive antenna and is modeled as \cite{jin2020spectral}
\begin{equation}
h_{j,i}^{(m)} = \sqrt{\beta_{j,i}^{(m)}}\left(
\sqrt{\frac{k_{j,i}^{(m)}}{k_{j,i}^{(m)}+1}}\,\bar{h}_{j,i}^{(m)}
+
\sqrt{\frac{1}{k_{j,i}^{(m)}+1}}\,\hat{h}_{j,i}^{(m)}
\right).
\end{equation}

In the above expression, $\beta_{j,i}^{(m)}$ denotes the large-scale fading coefficient, while the term in parentheses captures the small-scale fading. Specifically, $\bar{h}_{j,i}^{(m)}$ models the deterministic LoS component, and $\hat{h}_{j,i}^{(m)}$ represents the scattered non-line-of-sight (NLoS) component arising from the superposition of numerous multipath contributions. The parameter $k_{j,i}^{(m)}$ is the Rician $K$-factor, i.e., the power ratio between the LoS and NLoS components.

Consequently, the channel matrix between the $m$-th waveguide and the receiver can be compactly written as
\begin{equation}
\mathbf{H}_{m} =
\mathbf{B}_{m} \odot
\big(
\mathbf{K}_{{\mathrm{LoS}},m}\odot\mathbf{\bar{H}}_{m}
+
\mathbf{K}_{{\mathrm{NLoS}},m}\odot\mathbf{\widehat{H}}_{m}
\big),
\label{Hn}
\end{equation}
where the matrices $\mathbf{B}_{m}$, $\mathbf{K}_{{\mathrm{LoS}},m}$, $\mathbf{\bar{H}}_{m}$, $\mathbf{K}_{{\mathrm{NLoS}},m}$, and $\mathbf{\widehat{H}}_{m}$ collect, respectively, the element-wise values of $\sqrt{\beta_{j,i}^{(m)}}$, $\sqrt{\frac{k_{j,i}^{(m)}}{k_{j,i}^{(m)}+1}}$, $\bar{h}_{j,i}^{(m)}$, $\sqrt{\frac{1}{k_{j,i}^{(m)}+1}}$, and $\hat{h}_{j,i}^{(m)}$.

Finally, the overall MIMO channel matrix $\mathbf{H}$ of the PASM system is obtained by concatenating all sub-channel matrices associated with different waveguides:
\begin{equation}
\mathbf{H} = \big[\mathbf{H}_1,\ \mathbf{H}_2,\ \ldots,\ \mathbf{H}_{N_{\mathrm{wg}}}\big].
\label{H}
\end{equation}
This formulation enables the PASM model to jointly account for the geometric deployment of PAs and the underlying propagation characteristics, thereby providing a realistic basis for performance analysis and detector design.

\subsection{Signal Model}
In the proposed PASM system, the phase shifts are adjusted according to the positions of the activated PAs. Let $M_b$ and $M_p$ denote the modulation orders for the baseband and phase shifters, respectively. 
\begin{figure}[h]
\centerline{\includegraphics[width=0.9\linewidth]{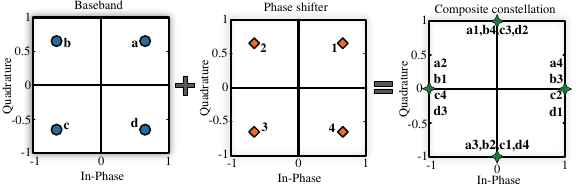}}
\caption{Schematic illustration of the composite constellation construction and the associated constellation ambiguity in the proposed PASM system.}
\label{xingzuo}
\end{figure}
The transmitter architecture and signal flow of the proposed PASM system are schematically illustrated in Fig. \ref{system}(a). As shown, the incoming bit stream is divided into two distinct streams. The first stream, comprising $N_{\mathrm{wg}}\log_{2}(M_b)$ bits, is mapped onto the corresponding APM symbols. The second stream, comprising $(N_t^\mathrm{ant}-N_{\mathrm{wg}}) \log_2 (M_p)$ bits, is modulated through the activated PA positions across waveguides. The constellations for baseband modulation, PA position selection, and phase modulation (via PA activation positions), as well as the antenna side, are shown in Fig. \ref{xingzuo}. It is important to note that we distinguish between the modulation orders for the baseband side and the PA position selection side (i.e., phase modulation) to account for the limited resolution of the PA position selection. In other words, the phase modulation order $M_p$ must be determined according to the minimum moving distance of PA, as it directly impacts the resolution of the phase shift that can be applied to each antenna element, while the baseband modulation order $M_b$ can be freely chosen. The antenna constellation diagram presented in Fig. \ref{xingzuo} illustrates that four different combinations of baseband and phase shift position selection constellations can result in the same antenna constellation, a phenomenon referred to as constellation ambiguity. To mitigate this issue, the initial antenna on each waveguide is anchored to a fixed reference position, typically corresponding to a zero phase shift. This configuration allows the first PA to serve as a phase reference, enabling the receiver to unambiguously distinguish the modulation of the baseband symbol from the spatial phase shifts carried by the subsequent PAs.  

After determining the modulation orders $M_b$ and $M_p$, the first modulation operation is performed at the baseband to obtain the temporary transmit vector $\mathbf{x}_b \in \mathbb{C}^{N_{\mathrm{wg}} \times 1}$, expressed as $\mathbf{x}_b = [s_1^b, s_2^b, \dots, s_{N_{\mathrm{wg}}}^b]^T$. Note that since all $N_a$ activated PAs are located on the same waveguide, they radiate the identical baseband symbol \cite{ding2025flexible}. Furthermore, we can generate the symbol vector at the waveguide level $\mathbf{x}_{\mathrm{wg}} \in \mathbb{C}^{N_t^{\mathrm{ant}} \times 1}$, which is given by
\begin{equation}
\mathbf{x}_{\mathrm{wg}}=\mathbf{E}\mathbf{x}_{b}=\left[\left\{s_{1}^{b},\ldots,s_{1}^{b}\right\},\ldots,\left\{s_{N_{\mathrm{wg}}}^{b},\ldots,s_{N_{\mathrm{wg}}}^{b}\right\}\right]^{\mathrm{T}},
\end{equation}
where $\mathbf{E} \in \mathbb{C}^{N_t^{\mathrm{ant}} \times N_{\mathrm{wg}}}$ is a sparse constant matrix for dimension extension, defined as $\mathbf{E} = \text{diag}\{\boldsymbol{1}, \boldsymbol{1}, \dots, \boldsymbol{1}\}$, with $\boldsymbol{1}$ being a $N_a \times 1$ all-ones column vector.

Furthermore, to elaborate on the physical mechanism underpinning this modulation, it is worth emphasizing that all PAs are located along their respective waveguides at specific distances from the feed points. The guided-wave propagation between the injection point $\mathbf{p}_m^{\mathrm{fp}}$ and the $i$-th activated PA at $\mathbf{p}_{m,i}^{\mathrm{tx}}$ inevitably introduces a deterministic phase shift in the radiated signal~\cite{ding2025flexible}. To capture this effect, we model the phase shift coefficient associated with the $i$-th PA on the $m$-th waveguide as
\begin{equation}
\alpha_i^{(m)} = \exp\!\left( -j \frac{2\pi}{\lambda_g}
\left\| \mathbf{p}_{m,i}^{\mathrm{tx}} - \mathbf{p}_m^{\mathrm{fp}} \right\|_2 \right),
\end{equation}
where $\lambda_g = \lambda / n_{\mathrm{eff}}$ denotes the guided wavelength in the dielectric waveguide, $\lambda = c / f_c$ is the free-space wavelength at carrier frequency $f_c$, $c$ is the speed of light, and $n_{\mathrm{eff}}$ is the effective refractive index of the waveguide \cite{pozar2021microwave}.

Finally, with the aid of the PA position-induced phase shifts, the transmit vector at the antenna side, $\mathbf{x}$, can be expressed as
\begin{equation}
\begin{aligned}
\mathbf{x} & =\mathbf{x}_{\mathrm{wg}}\odot\mathbf{a} \\
 & =\left[\underbrace{\left\{s_1^b,s_1^b\alpha_2^{(1)},\ldots,s_1^b\alpha_{N_a}^{(1)}\right\}}_{1\times N_a},\left\{s_2^b,s_2^b\alpha_2^{(2)},\ldots,s_2^b\alpha_{N_a}^{(2)}\right\},\ldots\right. \\
 & \left.\left\{s^{b}_{N_{\mathrm{wg}}},s^{b}_{N_{\mathrm{wg}}}\alpha_2^{(N_{\mathrm{wg}})},\ldots,s^{b}_{N_{\mathrm{wg}}}\alpha_{N_a}^{(N_\mathrm{wg})}\right\}\right]^{\mathrm{T}},
\end{aligned}
\end{equation}
where $\mathbf{a}$ is the phase shift symbol vector, denoted as
\begin{equation}
\begin{aligned}
\mathbf{a} & =\left[\left\{1,\alpha_2^{(1)},\ldots,\alpha_{N_a}^{(1)}\right\},\left\{1,\alpha_2^{(2)},\ldots,\alpha_{N_a}^{(2)}\right\},\ldots\right. \\
 & \left.\left\{1,\alpha_2^{(N_{\mathrm{wg}})},\ldots,\alpha_{N_a}^{(N_\mathrm{wg})}\right\}\right]^{\mathrm{T}}\\
 & = \left[ \mathbf{a}_1^T,\mathbf{a}_2^T\ldots, \mathbf{a}_{N_{\mathrm{wg}}}^T\right]^T.
\end{aligned}
\end{equation}

Finally, the spectral efficiency of the proposed PASM system can be readily calculated as
\begin{equation}
\eta=\underbrace{N_{\mathrm{wg}}\log_{2}M_b}_{\mathrm{APM~Part}}+\underbrace{(N_t^{\mathrm{ant}}-N_{\mathrm{wg}})\log_{2}M_p}_{\mathrm{PA~Position~Part}}\quad\mathrm{bits/s/Hz.}
\label{eta}
\end{equation}

Upon transmitting the symbol vector $\mathbf{x}$ over the fading channel, the received signal vector $\mathbf{y}\in\mathbb{C}^{N_{r}\times1}$ can be formulated as
\begin{equation}
\mathbf{y}=\sqrt{\delta}\mathbf{H}\mathbf{x}+\mathbf{n}=\sqrt{\delta}\mathbf{H}(\mathbf{x}_{\mathrm{wg}} \odot\mathbf{a})+\mathbf{n},
\end{equation}
where $\delta = \frac{P}{N_{\mathrm{wg}} N_a}$ is the normalized transmit power per activated antenna, with $P$ representing the overall transmit power, and $\mathbf{H}$ is the overall channel matrix as defined in \eqref{H}, and $\mathbf{n} \in \mathbb{C}^{N_r \times 1}$ is the noise vector, whose entries are assumed to be i.i.d. complex Gaussian random variables distributed as $\mathcal{CN}(0, N_0)$.

Consequently, the optimal ML detector for the PASM system jointly infers both the transmitted baseband symbols $\mathbf{x}_b$ and the phase shift symbols $\mathbf{a}$. Its decision rule can be formulated as
\begin{equation}
\left[\widehat{\mathbf{x}}_{b}^{\mathrm{ML}},\,\widehat{\mathbf{a}}\right]
    =\arg\min_{\mathbf{x}_{b}\in\mathcal{B}^{N_{\mathrm{wg}}},\,\mathbf{a}\in\mathcal{A}^{N_a}}
    \left\| \mathbf{y} - \sqrt{\delta}\,\mathbf{H}\mathbf{E}\left(\mathbf{x}_{b}\odot\mathbf{a}\right) \right\|_{2},
\end{equation}
where $\mathcal{B}^{N_{\mathrm{wg}}}$ and $\mathcal{A}^{N_a}$ denote the sets of all possible baseband symbol vectors and phase shift vectors, respectively.

\section{Vector Approximate Message Passing Detector Design}
In this section, we investigate the signal detection problem for the proposed PASM system. We commence by reviewing the optimal ML detector and the conventional linear detectors, analyzing their respective limitations in terms of computational complexity and error propagation. Motivated by these challenges, we then conceive a sophisticated VAMP detector tailored for the PASM system, which exploits the waveguide-structured prior to strike a compelling trade-off between performance and complexity.

\subsection{Conventional Linear Detector}
While conventional ML detection theoretically achieves optimal BER performance, its practical implementation is often hindered by prohibitive computational complexity. This complexity barrier arises from the exhaustive search required across the exponentially large solution space encompassing all legitimate combinations of modulated symbols and phase configurations. 

To circumvent this complexity barrier, classical linear detectors, such as zero-forcing (ZF) and minimum mean square error (MMSE) for the PASM system can be expressed as
\begin{equation}
\widehat{\mathbf{x}}^{\mathrm{ZF}}=\left(\sqrt{\delta}\mathbf{H}^{\mathrm{H}}\mathbf{H}\right)^{-1}\mathbf{H}^{\mathrm{H}}\mathbf{y}
\end{equation}
and
\begin{equation}
\mathbf{\widehat{x}}^{\mathrm{MMSE}}=\left(\sqrt{\delta}\mathbf{H}^{\mathrm{H}}\mathbf{H}+N_0\mathbf{I}\right)^{-1}\sqrt{\delta}\mathbf{H}^{\mathrm{H}}\mathbf{y}.
\end{equation}

After applying the above linear detectors, we obtain an estimate of the transmitted PASM symbol vector, denoted by $\widehat{\mathbf{x}} \in \mathbb{C}^{N_t^{\mathrm{ant}} \times 1}$. Based on $\widehat{\mathbf{x}}$, the estimates of the baseband symbol vector $\widehat{\mathbf{x}}_b$ and the phase vector $\widehat{\mathbf{a}}$ are recovered as
\begin{equation}
  \widehat{\mathbf{x}}_b
  = \big[\hat{x}(1),\, \hat{x}(1+N_a),\, \ldots,\, \hat{x}\big(1+(N_{\mathrm{wg}}-1)N_a\big)\big]^{\mathsf{T}},
  \label{eq:xb_est_lin}
\end{equation}
and
\begin{equation}
  \widehat{\mathbf{a}} = \widehat{\mathbf{x}} \oslash \big(\mathbf{E}\widehat{\mathbf{x}}_b\big),
  \label{eq:p_est_lin}
\end{equation}
where $\hat{x}(i)$ denotes the $i$-th element of $\widehat{\mathbf{x}}$.

However, this linear detection suffers from fundamental limitations. From a Bayesian perspective, linear estimators are suboptimal for PASM signals, as they relax the discrete constellation constraints to a continuous space, thereby discarding critical a priori information regarding the non-linear block-sparse structure inherent in $\mathbf{x} = \mathbf{x}_{\mathrm{wg}} \odot \mathbf{a}$. Furthermore, classical linear processing inevitably leads to noise enhancement or residual interference, degrading the effective SINR. Crucially, these estimation inaccuracies culminate in a severe error-propagation effect during the sequential recovery: as indicated by \eqref{eq:p_est_lin}, any decision error in the baseband symbol estimate $\widehat{\mathbf{x}}_b$ acts as a multiplicative distortion, directly contaminating the corresponding phase entries in $\widehat{\mathbf{a}}$.

To circumvent this limitation, we employ the VAMP algorithm for the PASM system. By exploiting the inherent structure of the PASM signal model and the statistical properties of the channel, the VAMP detector achieves substantially better performance than conventional linear detectors while maintaining polynomial-time computational complexity, thereby rendering the PASM system more viable for deployment.

\subsection{Factor Graph Representation}
\begin{figure}[htbp]
\centerline{\includegraphics[width=0.7\linewidth]{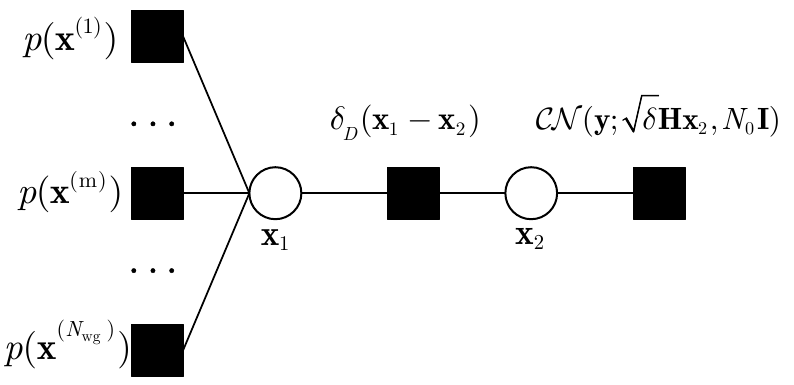}}
\caption{The factor graph corresponding to Eq. (\ref{x1x2}).}
\label{factor graph}
\end{figure}
To facilitate the derivation of our detection algorithm, we begin by formulating the factor graph representation of the PASM system. Following the standard procedure in \cite{rangan2019vector}, we first split the unknown signal $\mathbf{x}$ into two identical variables $\mathbf{x}_{1} = \mathbf{x}_{2}$, yielding the equivalent factorization
\begin{equation}
p( \mathbf{y},\mathbf{x}_{1}, \mathbf{x}_{2}) = p(\mathbf{x}_{1}) \delta_D(\mathbf{x}_{1} - \mathbf{x}_{2}) \mathcal{CN}(\mathbf{y} ; \sqrt{\delta}\mathbf{H}\mathbf{x}_{2}, N_0\mathbf{I}),
\label{x1x2}
\end{equation}
where $\delta_D(\cdot)$ is the Dirac delta distribution.

Unlike conventional AMP, which treats all signal elements as independent, the PASM system exhibits structured correlation that must be properly modeled. Specifically, since the $N_a$ elements on the same waveguide share the identical baseband symbol $s_m^b$ but with different phase shifts, the prior distribution of $\mathbf{x}$ should be organized by waveguide index $m$. This leads to a block-structured prior:
\begin{equation}
p(\mathbf{x}) = \prod_{m=1}^{N_{\mathrm{wg}}} p(\mathbf{x}^{(m)}),
\end{equation}
where $\mathbf{x}^{(m)} = s_m^b \odot \mathbf{a}_{m} \in \mathbb{C}^{N_a \times 1}$ represents the signal on the $m$-th waveguide. This waveguide-level factorization is visualized in Fig. \ref{factor graph}, where the prior constraint is applied at the waveguide level rather than at the individual antenna level.

Therefore, the probability density function (PDF) of $\mathbf{x}^{(m)}$ can be expressed as
\begin{equation}
p(\mathbf{x}^{(m)}) = \sum_{l=1}^{M_b M_p^{N_a-1}} \rho_l \delta_D(\mathbf{x}^{(m)} - \bar{\mathbf{x}}_l),
\end{equation}
where $\bar{\mathbf{x}}_l\in\mathcal{X}$, in which $\mathcal{X}$ denotes the set of all possible $\mathbf{x}^m$ with $|\mathcal{X}|=M_b M_p^{N_a-1}$, and $\rho_l$ represents the corresponding prior probability with $\sum_{l}\rho_{l}=1$.

The VAMP algorithm operates by passing messages on the factor graph according to three fundamental rules:

 1) \textit{Approximate Beliefs:} The approximate belief $b_{\mathrm{app}(\mathbf{x})}$ on variable node $\mathbf{x}$ is $\mathcal{CN}(\mathbf{x}; \widehat{\mathbf{x}}, \eta^{-1}\mathbf{I}_{\!N_t^{\mathrm{ant}}}\big))$, where $\widehat{\mathbf{x}} = \mathbb{E}[\mathbf{x} | b_{\mathrm{sp}}]$ and $\eta^{-1} = \langle \mathrm{diag}(\mathrm{Cov}[\mathbf{x} | b_{\mathrm{sp}}]) \rangle$ are the mean and average variance of the corresponding sum-product (SP) belief $b_{\mathrm{sp}}(\mathbf{x})\propto\prod_i\mu_{f_i\to\mathbf{x}}(\mathbf{x})$, that is, the normalized composite of all incoming messages incident upon the node.

 2) \textit{Variable-to-Factor Messages:} The message from variable node $\mathbf{x}$ to factor node $f_j$ is $\mu_{\mathbf{x} \to f_j}(\mathbf{x}) \propto b_{\mathrm{app}}(\mathbf{x}) / \mu_{f_j \to \mathbf{x}}(\mathbf{x})$.

 3) \textit{Factor-to-Variable Messages:} The message from factor node $f$ to variable node $\mathbf{x}$ is
\begin{equation}
\mu_{f \to \mathbf{x}}(\mathbf{x}) \propto \int f(\mathbf{x}, \{\mathbf{x}^{(j)}\}_{j \neq i}) \prod_{j \neq i} \mu_{\mathbf{x}^{(j)} \to f}(\mathbf{x}^{(j)}) d\mathbf{x}^{(j)}.
\end{equation}

\subsection{Waveguide-Level VAMP Algorithm}

Building upon the factor graph representation, we now derive the VAMP detector for the PASM system. The VAMP algorithm can be equivalently expressed using the linear MMSE (LMMSE) estimation framework, as shown in Algorithm 1. The algorithm iteratively exchanges information between two estimation modules: a waveguide-level denoiser and a linear MMSE estimator.
 
The key in our VAMP detector lies in the waveguide-level denoiser $g_1(\cdot, \gamma_{1k})$, which incorporates structural prior information. For the $m$-th waveguide, the posterior distribution combines the current estimate with the prior information and can be expressed as
\begin{equation}
p\big(\mathbf{x}^{(m)} = \bar{\mathbf{x}}_l \mid \mathbf{r}_{1k}^{(m)},\gamma_{1k}\big)
\propto \rho_l \prod_{i=1}^{N_a}
\mathcal{CN}\!\Big(r_{1k,i}^{(m)};\, [\bar{\mathbf{x}}_{l}^{(m)}]_i,\,\gamma_{1k}^{-1}\Big),
\end{equation}
where $\mathbf{r}_{1k}^{(m)} = \mathbf{x}^{(m)} + \mathbf{w}_{1k}, \mathbf{w}_{1k} \sim \mathcal{CN}\big(\mathbf{0},\,\gamma_{1k}^{-1}\mathbf{I}_{Na}\big)$ denotes the input estimate vector for the $m$-th waveguide at the $k$-th iteration, and $\gamma_{1k}$ is a precision parameter associated with the first estimation module at iteration $k$, which quantifies the reliability of $\mathbf{r}_{1k}^{(m)}$. 

Consequently, the waveguide-level denoiser is derived as
\begin{equation}
\mathbf{g}_1(\mathbf{r}_{1k}, \gamma_{1k}) = \widehat{\mathbf{x}}_{1k}
= \mathbb{E}\!\Big[\mathbf{x} \,\big|\, \mathbf{r}_{1k}, \gamma_{1k}\Big].
\end{equation}

For the $m$-th waveguide, this can be explicitly written as
\begin{equation}
\widehat{\mathbf{x}}_{1k}^{(m)}
= \mathbb{E}\!\big[\mathbf{x}^{(m)} \mid \mathbf{r}_{1k}^{(m)}, \gamma_{1k}\big]
= \sum_{l} w_l^{(m)}\, \bar{\mathbf{x}}_l,
\end{equation}
where $w_l$ can be expressed as 
\begin{equation}
w_l^{(m)}
= \frac{\rho_l \prod_{i=1}^{N_a}
  \mathcal{CN}\Big(r_{1k,i}^{(m)};\, [\bar{\mathbf{x}}_l^{(m)}]_i,\, \gamma_{1k}^{-1}\Big)}
 {\sum_{l'}^{{M_b M_p^{N_a-1}}} \rho_{l'} \prod_{i=1}^{N_a}
  \mathcal{CN}\Big(r_{1k,i}^{(m)};\, [\bar{\mathbf{x}}_{l'}^{(m)}]_i,\, \gamma_{1k}^{-1}\Big)},
\end{equation}
where $r_{1k,i}^m$ denotes the $i$-th element of the input of the $m$-th waveguide. This optimization jointly estimates the baseband symbol and phase shifts within each waveguide, the average divergence of this denoiser is given by
\begin{equation}
\begin{aligned}
\alpha_{1k} &= \langle \mathbf{g}_1'(\mathbf{r}_{1k}, \gamma_{1k}) \rangle= \gamma_{1k}\,\mathrm{Var}\!\Big[\mathbf{x}^{(m)} \,\Big|\, \mathbf{r}_{1k}^{(m)},\gamma_{1k}\Big]\\
& = \frac{1}{N_t^{\mathrm{ant}}} \sum_{m=1}^{N_{\mathrm{wg}}} \sum_{i=1}^{N_a} \frac{\partial \mathrm{g}_{1,i}^{(m)}}{\partial r_{1k,i}^{(m)}},
\end{aligned}
\end{equation}
where $\mathrm{g}_{1,i}^m$ denotes the $i$-th element of the output corresponding to the $m$-th waveguide. This divergence calculation captures the sensitivity of the waveguide-wise denoiser to its Gaussian input and is critical for the Onsager-type correction in VAMP, which mitigates correlation between iterations and promotes stable convergence.

The LMMSE estimator $\mathbf{g}_2(\mathbf{r}_{2k}, \gamma_{2k})$ in the PASM context is given by
\begin{equation}
\mathbf{g}_2(\mathbf{r}_{2k}, \gamma_{2k}) = \left(\gamma_w \mathbf{H}^H\mathbf{H} + \gamma_{2k}\mathbf{I}\right)^{-1} \left(\gamma_w \mathbf{H}^H\mathbf{y} + \gamma_{2k}\mathbf{r}_{2k}\right),
\end{equation}
which can be interpreted as the MMSE estimate of a random vector $\mathbf{x}_{2}$ under likelihood $\mathcal{CN}(\mathbf{y}; \sqrt{\delta}\mathbf{H}\mathbf{x}_{2}, \gamma_w^{-1}\mathbf{I})$ with $\gamma_w=1/N_0$ and prior $\mathbf{x}_{2} \sim \mathcal{N}(\mathbf{r}_{2k}, \gamma_{2k}^{-1}\mathbf{I})$, where $\gamma_{2k}$ is the precision of the Gaussian message from the prior-side module to the linear module at iteration $k$. Its divergence is
\begin{equation}
\langle \mathbf{g}_2'(\mathbf{r}_{2k}, \gamma_{2k}) \rangle = \frac{\gamma_{2k}}{N_t^{\mathrm{ant}}} \mathrm{Tr}\left[ \left(\gamma_w \mathbf{H}^H\mathbf{H} + \gamma_{2k}\mathbf{I}\right)^{-1} \right].
\end{equation}

Subsequently, in the final iteration of the VAMP algorithm, the denoiser computes the likelihood ratios (LLRs) for the baseband bits ($L_{b,j}$) and phase shift bits ($L_{p,k}$) as
\begin{align}
    L_{b,j}^{(m)} &= \ln \frac{\sum_{\mathbf{c}_l \in \mathcal{X}_{b,j}^{(1)}} p_{m,l}}{\sum_{\mathbf{c}_l \in \mathcal{X}_{b,j}^{(0)}} p_{m,l}}, \quad j = 1, \dots, \log_2 M_b, \\
    L_{p,k}^{(m)} &= \ln \frac{\sum_{\mathbf{c}_l \in \mathcal{X}_{p,k}^{(1)}} p_{m,l}}{\sum_{\mathbf{c}_l \in \mathcal{X}_{p,k}^{(0)}} p_{m,l}}, \quad k = 1, \dots, (N_a-1)\log_2 M_p,
\end{align}
where $\mathcal{X}_{b,j}^{(v)}$ $(\mathcal{X}_{p,k}^{(v)})$ represents the subset of $\mathcal{X}$ that the $j$-th baseband ($k$-th phase shift) bit is $v\in\{0,1\}$, and the $p_{m,l}=\frac{\exp\left( -\gamma_{1} \|\mathbf{r}_{1}^{(m)} - \mathbf{c}_l\|^2 \right)}{\sum_{j=1}^{|\mathcal{X}|} \exp\left( -\gamma_{1} \|\mathbf{r}_{1}^{(m)} - \mathbf{c}_j\|^2 \right)}$ denotes the posterior probability for each valid composite symbol vector $\mathbf{c}_l \in \mathcal{X}$ on the $m$-th waveguide, where $\mathbf{r}_{1}^{(m)}$ is the denoised observation and $\gamma_{1}$ is the precision. 

The complete algorithm proceeds as outlined in Algorithm 1, with the key difference from standard VAMP being the waveguide-structured denoiser. The algorithm maintains the robust convergence properties of VAMP while explicitly modeling the correlation within each waveguide.

\subsection{Implementation Details and Complexity Analysis}
For practical implementation with finite-dimensional matrices, we incorporate several enhancements to improve robustness and convergence. First, we clip the precision parameters $\gamma_{1k}$ and $\gamma_{2k}$ to a positive interval $[\gamma_{\min}, \gamma_{\max}]$. While uncommon in well-conditioned systems, the denoiser's average divergence $\alpha_{1k}$ may occasionally become negative, which would lead to invalid precision estimates if not accounted for. For the numerical results presented in Section V, we used $\gamma_{\min} = 1 \times 10^{-11}$, $\gamma_{\max} = 1 \times 10^{11}$.

Second, we incorporate damping to enhance convergence stability, i.e.,
\begin{align}
\widehat{\mathbf{x}}_{1,k+1} &= \beta \mathbf{g}_1(\mathbf{r}_{1k}, \gamma_{1k}) + (1-\beta)\widehat{\mathbf{x}}_{1k}, \\
\widehat{\mathbf{x}}_{2,k+1} &= \beta \mathbf{g}_2(\mathbf{r}_{2k}, \gamma_{2k}) + (1-\beta)\widehat{\mathbf{x}}_{2k},
\end{align}
where $\beta \in (0,1]$ is a damping parameter. For our simulations, we used $\beta = 0.6$.

Third, rather than requiring a fixed number of iterations, we terminate the algorithm when the normalized difference $\|\mathbf{r}_{1,k+1} - \mathbf{r}_{1k}\|/\|\mathbf{r}_{1,k+1}\|$ falls below a tolerance $\epsilon$. For our simulations, we used $\epsilon = 1 \times 10^{-4}$.

As for the computational complexity, the optimal ML detector needs to exhaustively search over all feasible transmit vectors, whose cardinality is 
$M_b^{N_{\mathrm{wg}}} M_p^{N_t^{\mathrm{ant}}-N_{\mathrm{wg}}}$.  
For each candidate vector, this leads to an exponential complexity on the order of
$\mathcal{O}\big(N_r N_t^{\mathrm{ant}} M_b^{N_{\mathrm{wg}}} M_p^{N_t^{\mathrm{ant}}-N_{\mathrm{wg}}}\big)$, which is prohibitive even for moderate dimensions.

For the proposed detector, the per-iteration complexity is dominated by the linear module and the waveguide-structured denoiser. 
The linear module performs one LMMSE estimation, which requires 
$\mathcal{O}\big((N_t^{\mathrm{ant}})^3\big)$ operations. 
The prior-side module jointly processes the $N_a$ PAs on each of the $N_{\mathrm{wg}}$ waveguides using a discrete composite-constellation prior, resulting in an additional complexity of approximately 
\begin{figure}[t]
\begin{algorithm}[H]
    \caption{Proposed Detector for PASM System}\label{alg:cap}
    \begin{algorithmic}[1]
    \Require Received signal $\mathbf{y}$, channel matrix $\mathbf{H}$, noise precision $\gamma_w$, LMMSE estimator $\mathbf{g}_2(\cdot,\cdot)$, denoiser $\mathbf{g}_1(\cdot,\gamma_{1k})$, maximum iterations $T_{\max}$.
\Ensure Estimates $\widehat{\mathbf{x}}_b$ and $\widehat{\mathbf{a}}$.
\State \textbf{Initialize:} $\mathbf{r}_{10}$, $\gamma_{10} \geq 0$;
\For{$k = 0,1$ to $T_{\max}$}
\State // \textit{Denoising}
\State $\widehat{\mathbf{x}}_{1k} = \mathbf{g}_1(\mathbf{r}_{1k}, \gamma_{1k})$;
\State $\alpha_{1k} = \langle \mathbf{g}_1'(\mathbf{r}_{1k}, \gamma_{1k}) \rangle$;
\State $\eta_{1k} = \gamma_{1k}/\alpha_{1k}$;
\State $\gamma_{2k} = \eta_{1k} - \gamma_{1k}$;
\State $\mathbf{r}_{2k} = (\eta_{1k}\widehat{\mathbf{x}}_{1k} - \gamma_{1k}\mathbf{r}_{1k})/\gamma_{2k}$;
\State // \textit{LMMSE estimation}
\State $\widehat{\mathbf{x}}_{2k} = \mathbf{g}_2(\mathbf{r}_{2k}, \gamma_{2k})$;
\State $\alpha_{2k} = \langle \mathbf{g}_2'(\mathbf{r}_{2k}, \gamma_{2k}) \rangle$;
\State $\eta_{2k} = \gamma_{2k}/\alpha_{2k}$;
\State $\gamma_{1,k+1} = \eta_{2k} - \gamma_{2k}$;
\State $\mathbf{r}_{1,k+1} = (\eta_{2k}\widehat{\mathbf{x}}_{2k} - \gamma_{2k}\mathbf{r}_{2k})/\gamma_{1,k+1}$;
\EndFor
\For{$m = 1$ to $N_{\mathrm{wg}}$}
\State Calculate $\{p_{m,l}\}$;
\State Calculate $\{L_{b,j}^{(m)}\}$, $\{L_{p,k}^{(m)}\}$;
\EndFor
\State \textbf{return} $\{L_{b,j}^{(m)}\}$, $\{L_{p,k}^{(m)}\}$.
\end{algorithmic}
\end{algorithm}
\end{figure}
$\mathcal{O}\big(N_{\mathrm{wg}} N_a M_b M_p^{N_a-1}\big)$ per iteration. 
Hence, for $T_{\max}$ iterations, the overall complexity of the proposed VAMP detector scales polynomially as
$\mathcal{O}\big((N_t^{\mathrm{ant}})^3\big)$, which is vastly lower than the exponential complexity of ML detection.

\section{Performance Analysis of the PASM System}
In this section, the theoretical BER performance of the proposed PASM system is derived under the ML detection. An upper bound on the BER can be established by invoking the well-known union bound \cite{simon2004digital}, which is expressed as
\begin{equation}
P_b\leq\frac{1}{2^\eta}\sum_{i=1}^{2^\eta}\sum_{j=1}^{2^\eta}\frac{P(\mathbf{x}_i\to\mathbf{x}_j)n_{i,j}}{\eta},
\end{equation}
where $\eta$ denotes the spectral efficiency as defined in \eqref{eta}. The term $P(\mathbf{x}_i \to \mathbf{x}_j)$ represents the pairwise error probability (PEP), which indicates the likelihood of erroneously detecting the vector $\mathbf{x}_j$ when $\mathbf{x}_i$ was transmitted. Furthermore, $n_{i,j}$ quantifies the number of bits in error between the information sequences corresponding to $\mathbf{x}_i$ and $\mathbf{x}_j$. The detailed derivation of the PEP is provided in the following.

The conditional PEP under the channel $\mathbf{H}$ can be expressed as
\begin{equation}
    \begin{aligned}
&P\left(\mathbf{x}_{i}\to\mathbf{x}_{j}\right|\mathbf{H})  \\
& =P\left(\left\|\mathbf{y}-\sqrt{\delta}\mathbf{H}\mathbf{x}_{j}\right\|_{2}^{2}\leq\left\|\mathbf{y}-\sqrt{\delta}\mathbf{H}\mathbf{x}_{i}\right\|_{2}^{2}\right) \\
 & =P\left(\left\|\sqrt{\delta}\mathbf{H}\left(\mathbf{x}_{i}-\mathbf{x}_{j}\right)+\mathbf{n}\right\|_{2}^{2}\leq\left\|\mathbf{n}\right\|_{2}^{2}\right) \\
 & =Q\left(\sqrt{\frac{\delta\left\|\mathbf{H}\mathbf{\Psi}\right\|_{2}^{2}}{2N_0}}\right), 
\end{aligned}
\label{Eq conpep}
\end{equation}
where $\mathbf{\Psi}=\mathbf{x}_i-\mathbf{x}_j$ and $Q(x)=\frac{1}{\pi}\int_{0}^{\pi/2}\exp\left(-\frac{x^{2}}{2\mathrm{sin}^{2}\theta}\right)d\theta$ \cite{kalialakis2001digital} is the Gaussian Q-function. 

As indicated by the PEP analysis in \eqref{Eq conpep}, the achievable BER performance fundamentally depends on the received SNR, which is itself affected by the large-scale fading conditions. Specifically, since the large-scale channel gain, denoted by $\beta(d)$, is a monotonically decreasing function of the propagation distance $d$, maximizing the received SNR is equivalent to minimizing the physical distance between the transmitting PAs and the receiver.

Consequently, for each waveguide $m \in \{1,\ldots,N_{\mathrm{wg}}\}$, the transmitter orchestrates the activation of the PAs such that they are positioned as close to the target user as possible. Let $\mathcal{P}_m$ denote the set of all candidate PA positions on the $m$-th waveguide. The optimal activation position index $a^\star$ for a given symbol interval is selected according to the following criterion:
\begin{equation}
\begin{aligned}
a^\star &= \arg\max_{a \in \mathcal{P}_m} \beta\big(\|\mathbf{p}_n^{\mathrm{r}} - \mathbf{p}_{m,a}^{\mathrm{tx}}\|\big)\\
      & = \arg\min_{a \in \mathcal{P}_m} \big\|\mathbf{p}_n^\mathrm{r} - \mathbf{p}_{m,a}^\mathrm{tx}\big\|,
\end{aligned}
\label{eq:pa_selection}
\end{equation}
where $\mathbf{p}_{m,a}^{\mathrm{tx}}$ represents the $a$-th candidate PA position on the $m$-th waveguide.

By averaging (\ref{Eq conpep}) over the legitimate range of the random variable $\gamma=\left\|\mathbf{H}\mathbf{\Psi}\right\|_{2}^{2}$, the unconditional PEP can be expressed by
\begin{equation}
\begin{aligned}
&P\left(\mathbf{x}_{i}\to\mathbf{x}_{j}\right) \\
 & =\frac{1}{\pi}\int_{0}^{\pi/2}\int_{0}^{+\infty}\exp\left(-\frac{\delta\gamma}{4N_0\mathrm{sin}^{2}\theta}\right)p_{\gamma}\left(\gamma\right)d\gamma d\theta \\
 &=\frac{1}{\pi}\int_{0}^{\pi/2}M_{\gamma}\left(-\frac{\delta}{4N_0\mathrm{sin}^{2}\theta}\right)d\theta,
\end{aligned}
\label{Eq Pxixj}
\end{equation}
where $M_{\gamma}\left(a\right)=\int_{0}^{+\infty}e^{a\gamma}p_{\gamma}\left(\gamma\right)d\gamma$ represents the moment generating function (MGF) of $\gamma$ which can be further written as 
\begin{equation}
    \begin{aligned}
\gamma & =\left\|\mathbf{H}\mathbf{\Psi}\right\|_{2}^{2} \\
 & =\underbrace{\mathrm{vec}\left(\mathbf{H}^{H}\right)^{H}}_{\mathbf{u}^{H}}\underbrace{\left(\mathbf{I}_{N_{r}}\otimes\mathbf{\Psi}\mathbf{\Psi}^{H}\right)}_{\mathbf{Q}}\underbrace{\mathrm{vec}\left(\mathbf{H}^{H}\right)}_{\mathbf{u}} \\
 &=\mathbf{u}^{H}\mathbf{Q}\mathbf{u}.
\end{aligned}
\label{gamma}
\end{equation}

According to the result on the characteristic function of Hermitian quadratic forms in complex Gaussian random variables given in \cite{turin1960characteristic}, the MGF of $\gamma$ can be expressed as
\begin{equation}
    M_{\gamma}\left(a\right)=\frac{\exp\left(a\times\bar{\mathbf{u}}^{H}\mathbf{Q}\left(\mathbf{I}-a\mathbf{C}_{\mathbf{u}}\mathbf{Q}\right)^{-1}\bar{\mathbf{u}}\right)}{\left|\mathbf{I}-a\mathbf{C}_{\mathbf{u}}\mathbf{Q}\right|},
    \label{Eq Mgamma}
\end{equation}
where $\bar{\mathbf{u}}=\mathbb{E}\left\{\mathbf{u}\right\}$ is the vector of the means and $\mathbf{C}_{\mathbf{u}}=\operatorname{E}\left\{\left(\mathbf{u}-\mathbf{\bar{u}}\right)\left(\mathbf{u}-\mathbf{\bar{u}}\right)^{H}\right\}$ is the covariance matrix. By plugging (\ref{Eq Mgamma}) into (\ref{Eq Pxixj}), we arrive at (\ref{new pxixj}), shown at the top of the next page.

Now, we consider the derivation of $\bar{\mathbf{u}}$ and $\mathbf{C}_{\mathbf{u}}$. Substituting (\ref{Hn}) and (\ref{H}) into (\ref{gamma}) yields (\ref{u}). In (\ref{u}), $\mathbf{K}_{N_{r}N_{t}^\mathrm{ant}}$ is the commutation matrix defined as
\setcounter{equation}{37}
\begin{equation}
\mathbf{K}_{N_{r}N_{t}^\mathrm{ant}}\triangleq\sum_{j=1}^{N_{t}^\mathrm{ant}}\left(\mathbf{e}_{j}^{T}\otimes\mathbf{I}_{N_{r}}\otimes\mathbf{e}_{j}\right),
    \label{KNrNt}
\end{equation}
where $\mathbf{e}_j$ is the $j$-th column vector of the $N_r \times {N_t^\mathrm{ant}}$ identity matrix.

Thus, $\bar{\mathbf{u}}$ can be formulated as (\ref{ubar}) and $\mathbf{C}_{\mathbf{u}}$ is given by
\setcounter{equation}{39}
\begin{equation}
\begin{aligned}
\mathbf{C}_{\mathbf{u}} & =\mathbb{E}\left\{\mathrm{vec}\left(\mathbf{H}^{H}\right)\mathrm{vec}\left(\mathbf{H}^{H}\right)^{H}\right\}-\mathbf{\bar{u}}\mathbf{\bar{u}}^{H} \\
 & =\mathbf{K}_{N_{r}N_{t}^\mathrm{ant}}\cdot\mathrm{blkdiag}(\mathbf{C}_{\mathbf{u},1},\ldots,\mathbf{C}_{\mathbf{u},N_{\mathrm{wg}}})\cdot\mathbf{K}_{N_{r}N_{t}^\mathrm{ant}}^{T},
\end{aligned}
    \label{Cu}
\end{equation}
where $\mathbf{C}_{\mathbf{u},m}$ is expressed as (\ref{c(u,n)}).

There is no closed-form of (\ref{new pxixj}) so it can be evaluated numerically. Nevertheless, with the aid of the approximation of the Q-function in \cite{chiani2002improved} which can be expressed as 
\setcounter{equation}{41}
\begin{equation}
    Q\left(x\right)\approx\frac{1}{12}e^{-\frac{x^{2}}{2}}+\frac{1}{4}e^{-\frac{2x^{2}}{3}},x\geq0,
\end{equation}

the PEP can be approximated as
\begin{equation}
    \begin{aligned}
 & P\left(\mathbf{x}_{i}\to\mathbf{x}_{j}\right) \\
 & \approx\int_{0}^{\infty}\left(\frac{1}{12}e^{-\frac{\delta\gamma}{4N_{0}}}+\frac{1}{4}e^{-\frac{\delta\gamma}{3N_{0}}}\right)f_{\gamma}\left(\gamma\right)d\gamma \\
 & =  \frac{1}{12}M_{\gamma}\left(-\frac{\delta}{4N_{0}}\right)+\frac{1}{4}M_{\gamma}\left(-\frac{\delta}{3N_{0}}\right) \\
 & =  \frac{1}{12}\frac{\exp\left(-\frac{\delta}{4N_{0}}\bar{\mathbf{u}}^{H}\mathbf{Q}\left(\mathbf{I}+\frac{\delta}{4N_{0}}\mathbf{C}_{\mathbf{u}}\mathbf{Q}\right)^{-1}\bar{\mathbf{u}}\right)}{\left|\mathbf{I}+\frac{\delta}{4N_{0}}\mathbf{C}_{\mathbf{u}}\mathbf{Q}\right|} \\
 &  +\frac{1}{4}\frac{\exp\left(-\frac{\delta}{3N_{0}}\bar{\mathbf{u}}^{H}\mathbf{Q}_{i,j}\left(\mathbf{I}+\frac{\delta}{3N_{0}}\mathbf{C}_{\mathbf{u}}\mathbf{Q}\right)^{-1}\bar{\mathbf{u}}\right)}{\left|\mathbf{I}+\frac{\delta}{3N_{0}}\mathbf{C}_{\mathbf{u}}\mathbf{Q}\right|}.
\end{aligned}
\end{equation}

\begin{figure}[htbp]
		\normalsize
		\setcounter{equation}{35}
		\begin{equation}\label{new pxixj}
P\left(\mathbf{x}_{i}\to\mathbf{x}_{j}\right)=\frac{1}{\pi}\int_{0}^{\pi/2}\frac{\exp\left\{-\frac{\delta}{4N_{0}\sin^{2}\theta}\mathbf{\bar{u}}^{H}\mathbf{Q}_{i,j}\left(\mathbf{I}+\frac{\delta}{4N_{0}\sin^{2}\theta}\mathbf{C}_{\mathbf{u}}\mathbf{Q}\right)^{-1}\mathbf{\bar{u}}\right\}}{\left|\mathbf{I}+\frac{\delta}{4N_{0}\sin^{2}\theta}\mathbf{C}_{\mathbf{u}}\mathbf{Q}\right|}d\theta,
		\end{equation}
		\hrulefill
	\end{figure}

    \begin{figure}[htbp]
		\normalsize
		\setcounter{equation}{36}
		\begin{equation}\label{u}
\left.\mathbf{u}=\mathbf{K}_{N_{r}N_{t}^\mathrm{ant}}\left[\begin{array}{c}\mathrm{vec}\left(\mathbf{B}_1^{*}\right)\odot\left[\mathrm{vec}\left(\mathbf{K}_{\mathrm{LoS},1}^{*}\right)\odot\mathrm{vec}\left(\mathbf{\bar{H}}_1^{*}\right)+\mathrm{vec}\left(\mathbf{K}_{\mathrm{NLoS},1}^{*}\right)\odot\mathrm{vec}\left(\mathbf{\widehat{H}}_1^{*}\right)\right]\\ \mathrm{vec}\left(\mathbf{B}_1^{*}\right)\odot\left[\mathrm{vec}\left(\mathbf{K}_{\mathrm{LoS},2}^{*}\right)\odot\mathrm{vec}\left(\mathbf{\bar{H}}_2^{*}\right)+\mathrm{vec}\left(\mathbf{K}_{\mathrm{NLoS},2}^{*}\right)\odot\mathrm{vec}\left(\mathbf{\widehat{H}}_2^{*}\right)\right]\\ \vdots\\ \mathrm{vec}\left(\mathbf{B}_{N_{\mathrm{wg}}}^{*}\right)\odot\left[\mathrm{vec}\left(\mathbf{K}_{\mathrm{LoS},N_{\mathrm{wg}}}^{*}\right)\odot\mathrm{vec}\left(\mathbf{\bar{H}}_{N_{\mathrm{wg}}}^{*}\right)+\mathrm{vec}\left(\mathbf{K}_{\mathrm{NLoS},N_{\mathrm{wg}}}^{*}\right)\odot\mathrm{vec}\left(\mathbf{\widehat{H}}_{N_{\mathrm{wg}}}^{*}\right)\right]\end{array}\right.\right],
		\end{equation}
		\hrulefill
	\end{figure}

    \begin{figure}[htbp]
		\normalsize
		\setcounter{equation}{38}
		\begin{equation}\label{ubar}
\left.\bar{\mathbf{u}}=\mathbb{E}\left\{\operatorname{vec}\left(\mathbf{H}^{H}\right)\right\}=\mathbf{K}_{N_{r}N_{t}^\mathrm{ant}}\left[
\begin{array}
{c}\operatorname{vec}\left(\mathbf{B}_{1}^{*}\right)\odot\operatorname{vec}\left(\mathbf{K}_{\mathrm{LoS},1}^{*}\right)\odot\operatorname{vec}\left(\mathbf{\widehat{H}}_{1}^{*}\right)\\
\operatorname{vec}\left(\mathbf{B}_{2}^{*}\right)\odot\operatorname{vec}\left(\mathbf{K}_{\mathrm{LoS},2}^{*}\right)\odot\operatorname{vec}\left(\mathbf{\widehat{H}}_{2}^{*}\right) \\
\vdots \\
\operatorname{vec}\left(\mathbf{B}_{N_{\mathrm{wg}}}^{*}\right)\odot\operatorname{vec}\left(\mathbf{K}_{\mathrm{LoS},N_{\mathrm{wg}}}^{*}\right)\odot\operatorname{vec}\left(\mathbf{\widehat{H}}_{N_{\mathrm{wg}}}^{*}\right)
\end{array}\right.\right],
		\end{equation}
		\hrulefill
	\end{figure}

\begin{figure}[htbp]
		\normalsize
		\setcounter{equation}{40}
		\begin{equation}\label{c(u,n)}
\mathbf{C}_{\mathbf{u},m}=\mathrm{vec}\left(\mathbf{B}_{n}^{*}\right)\mathrm{vec}(\mathbf{B}_{(m)}^{*})^{H}\odot\mathrm{vec}\left(\mathbf{K}_{\mathrm{LoS},m}^{*}\right)\mathrm{vec}\left(\mathbf{K}_{\mathrm{LoS},m}^{*}\right)^{H}.
		\end{equation}
		\hrulefill
	\end{figure}

\section{Simulation Results}
This section presents the simulated performance of the proposed PASM architecture and corroborates the analytical upper bound of its BER performance.
\subsection{Simulation Environment and Parameters}
For clarity of exposition, we consider a square service region with side length $D = 500$~$\mathrm{m}$. The system operates at a carrier frequency of $f_c = 3$~GHz, with a noise power of $N_0 = -90$~dBm and an effective refractive index of the dielectric waveguide set to $n_{\mathrm{eff}} = 1.4$. As illustrated in Fig. \ref{system}(b), the receiver is positioned at $(400~\mathrm{m},\,50~\mathrm{m},\,1.5~\mathrm{m})$, with its ULA aligned along the $x$-axis. The antenna elements are spaced by half a wavelength ($\lambda/2$) to mitigate mutual coupling effects. The configurations of the proposed PASM system and the conventional PSSM system are detailed as follows.

1) \textit{Proposed PASM:} As illustrated in Fig. \ref{system}(b), the transmitter is positioned at the origin $(0~\mathrm{m}, 0~\mathrm{m})$ of the square service region at a height of $12.5~\mathrm{m}$. The system employs $N_{\mathrm{wg}}$ dielectric waveguides deployed parallel to the $x$-axis, with their positions equidistantly distributed along the $y$-axis. For each waveguide, candidate positions for the PAs are established in the vicinity of the optimal activation point closest to the user, with adjacent elements separated by $\lambda/2$ intervals. This spatial configuration mitigates large-scale fading and enhances the received SNR, as analytically justified in Section IV. Table I summarizes the system configurations considered in this section.

\begin{table}[htbp]
\centering
\renewcommand{\arraystretch}{1.5} 
\caption{Simulation Parameter Configuration for PASM Systems.}
\begin{tabular}{    >{\centering\arraybackslash}p{1.1cm}
    >{\centering\arraybackslash}p{0.6cm}
    >{\centering\arraybackslash}p{0.6cm}
    >{\centering\arraybackslash}p{0.6cm}
    >{\centering\arraybackslash}p{0.6cm}
    >{\centering\arraybackslash}p{0.6cm}
    >{\centering\arraybackslash}p{1.2cm}}
\hline
 & $N_t^\mathrm{ant}$  & $N_r$& $N_\mathrm{wg}$ & $M_b$ & $M_p$ & $P_t$ (dBm) \\
\hline
Fig.~4  & 2 & -- & 1 & -- & 2 & -- \\
Fig.~5  & 2 & 2 & 1 & -- & 4 & -- \\
Fig.~6 (a)  & 4 & 4 & 1 & 16 & 4 & -- \\
Fig.~6 (b) & 4 & 6 & 1 & 16 & 4 & --\\
Fig.~7  & 4 & 4 & 2 & -- & -- & --\\
Fig.~8  & 4 & 4 & 1 & -- & -- & 25 \\
\hline
\end{tabular}
\end{table}

Regarding the channel parameters, the LoS component gain is normalized to unity. In accordance with the 3GPP channel model \cite{network20063rd} and the guidelines in \cite{ozdogan2018cell,ozdogan2019massive}, the probability of a LoS component existing between the $i$-th active PA on the $m$-th waveguide and the $n$-th receive antenna is predominantly governed by the distance $d_{j,i}^m$. More specifically, in typical urban scenarios, this probability can be expressed as 
\setcounter{equation}{43}
\begin{equation}
    P(\mathrm{LoS})=
\begin{cases}
1-\frac{d_{j,i}^{(m)}}{300}, & 0<d_{j,i}^{(m)}<300\mathrm{(m)}, \\
0, & d_{j,i}^{(m)}\geq300\mathrm{(m)}. 
\end{cases}
\end{equation}

Accordingly, the Rician factor can be calculated as \cite{ozdogan2018cell
,ozdogan2019massive
}
\begin{equation}
    K_{j,i}^{(m)}=
\begin{cases}
10^{1.3-0.003d_{j,i}^{(m)}}, & \text{if}~P(\mathrm{LoS})>0, \\
0, & \text{if}~P(\mathrm{LoS})=0. 
\end{cases}
\end{equation}

For the path-loss characterization, we adopt the COST 231 Walfisch-Ikegami model, whereby the corresponding large-scale fading coefficients (in dB) are computed according to \cite{ozdogan2018cell
,ozdogan2019massive}
\begin{equation}
    \beta_{j,i}^{(m)}=
\begin{cases}
-30.18-26\log_{10}(d_{j,i}^{(m)})+F_{j,i}^{(m)}, & K_{j,i}^{(m)}\neq0, \\
-34.53-38\log_{10}(d_{j,i}^{(m)})+F_{j,i}^{(m)}, & K_{j,i}^{(m)}=0. 
\end{cases}
\end{equation}
The correlated shadow fading coefficient is modeled as 
\begin{equation}
    F_{j,i}^{(m)}=\sqrt{\xi}e_i^{(m)}+\sqrt{1-\xi}b_j,
\end{equation}
where $\xi=0.5$, $e_i^{(m)}\sim\mathcal{N}\left(0,\sigma^2\right)$, $b_j\sim\mathcal{N}\left(0,\sigma^2\right)$ and $\sigma=8$. The covariance function of $e_i^{(m)}$ and $b_j$ can be expressed as 
\begin{equation}
    \mathbb{E}\{e_i^{(m)}e_{i^{\prime}}^{{(m)}^{\prime}}\}=2^{-\frac{d_\mathrm{PA}(m,i;m^{\prime},i^{\prime})}{d_\mathrm{decorr}}},\quad\mathbb{E}\{b_jb_{j^{\prime}}\}=2^{-\frac{d_\mathrm{r}(j,j^{\prime})}{d_\mathrm{decorr}}},
\end{equation}
where $d_{\mathrm{PA}}(m,i;\, m',i')$ denotes the geographical separation between the $i$-th activated PA on the $m$-th waveguide and the $i'$-th activated PA on the $m'$-th waveguide, while $d_{\mathrm{r}}(j,j')$ represents the distance between the $j$-th and $j'$-th receive antennas. The parameter $d_{\mathrm{decorr}}$ denotes the decorrelation distance, which is environment dependent and typically chosen as $d_{\mathrm{decorr}} = 100~\mathrm{m}$.

 2) \textit{Conventional PSSM System:} The transmitter is positioned at the center of the square coverage area illustrated in Fig. \ref{system}(b), elevated at the height $d_h=12.5~\mathrm{m}$. Its antenna array is oriented along the $x$-axis, with adjacent elements spaced at half the wavelength.
 
\subsection{Results and Discussions}
\begin{figure}[htbp]
\centerline{\includegraphics[width=0.7\linewidth]{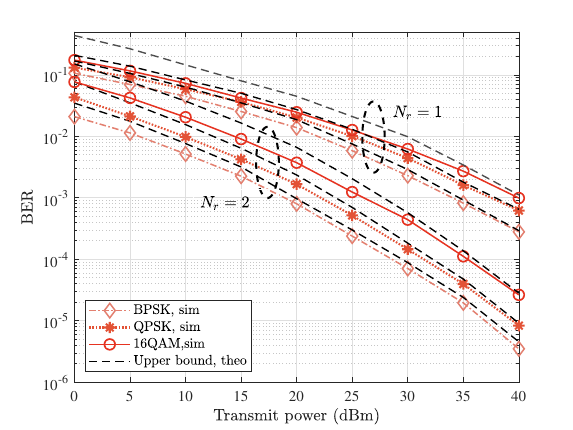}}
\caption{BER performance of the PASM system with different modulation orders and receive antenna numbers.}
\label{MLbounder}
\end{figure}
Fig. \ref{MLbounder} illustrates the BER performance of the proposed PASM system for different modulation orders, with the number of receive antennas set to $N_r=1$ and $N_r=2$. It can be readily observed that the simulated results exhibit an excellent agreement with the analytically derived upper bound, particularly in the high-SNR regime, thereby confirming the accuracy of the theoretical analysis. Furthermore, increasing the number of receive antennas yields a substantial performance improvement, indicating that enhanced receive diversity effectively mitigates the detrimental effects of fading.

\begin{figure}[htbp]
\centerline{\makebox[\linewidth]{\includegraphics[width=0.7\linewidth]{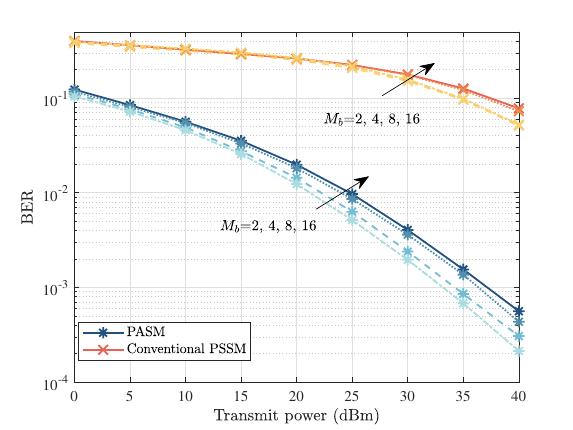}}}
\caption{BER performance of the PASM and conventional PSSM system.}
\label{vesus}
\end{figure}

To provide further insight into the performance gain brought by the proposed architecture, Fig.~\ref{vesus} compares the BER of the PASM system with that of the conventional PSSM scheme. For a fair comparison, both systems are configured with the same number of RF chains, transmit/receive antennas, and modulation orders so that they achieve the same spectral efficiency. In particular, for a spectral efficiency of $4$~bits/s/Hz, i.e., $(N_t^\mathrm{ant}, N_r, N_\mathrm{wg}) = (2,2,1)$, the PASM system attains approximately a $35$~dB SNR gain at a BER of $10^{-1}$ relative to PSSM. This performance improvement mainly stems from the flexible placement of the serving waveguide and active PAs close to the user, which establishes a strong LoS link, significantly mitigates large-scale path loss, and yields a higher received SNR than a fixed-array PSSM transmitter.

\begin{figure}[htbp]
    \centering
    
    \makebox[\linewidth]{%
        \subfloat[\label{1a}]{
            \includegraphics[width=0.7\linewidth]{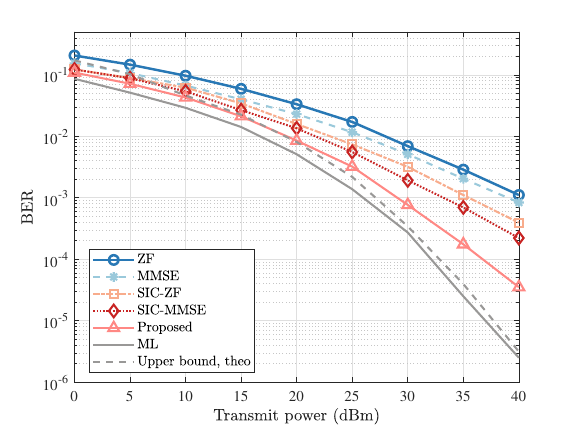}}
    }

    \makebox[\linewidth]{%
        \subfloat[\label{1b}]{
            \includegraphics[width=0.7\linewidth]{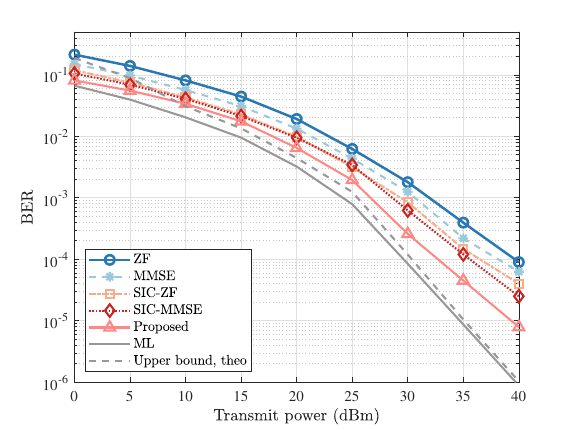}}
    }

    \caption{BER performance results of the PASM system using different detectors for (a) $N_t^\mathrm{ant}=4$, $N_r=4$, $M_b=16$, $M_p=4$ and (b) $N_t^\mathrm{ant}=4$, $N_r=6$, $M_b=16$, $M_p=4$.}
    \label{different detectors}
\end{figure}

Fig.~\ref{different detectors} depicts the BER performance of the PASM system with different detection algorithms. For benchmarking, we consider the low-complexity linear detectors, together with their successive-interference-cancellation (SIC) enhanced variants, i.e., SIC--ZF and SIC--MMSE. In addition, the ML detector is included as the optimal performance reference. In Fig.~\ref{different detectors}(a), for the configuration $(N_t^\mathrm{ant},N_r,M_b,M_p)=(4,4,16,4)$, the linear detectors exhibit the worst performance over the entire transmit-power range, while the SIC-based schemes provide a noticeable gain by partially cancelling inter-stream interference. The proposed VAMP-based detector further improves the BER compared with all linear and SIC baselines, especially in the medium-to-high SNR region, yet there remains a discernible performance gap with respect to the ML detector. Specifically, in Fig. \ref{different detectors}(a), the performance gap between the proposed algorithm and ML is approximately 2.9 dB at the BER of $10^{-3}$. A similar trend is observed in Fig.~\ref{different detectors}(b), where increasing the number of receive antennas uniformly enhances the performance of all schemes. In both scenarios, the proposed detector consistently outperforms the linear and SIC-based detectors, demonstrating the benefit of exploiting the PASM prior structure, while ML detection still provides the best performance at the expense of very high complexity.

\begin{figure}[htbp]
\centerline{\includegraphics[width=0.7\linewidth]{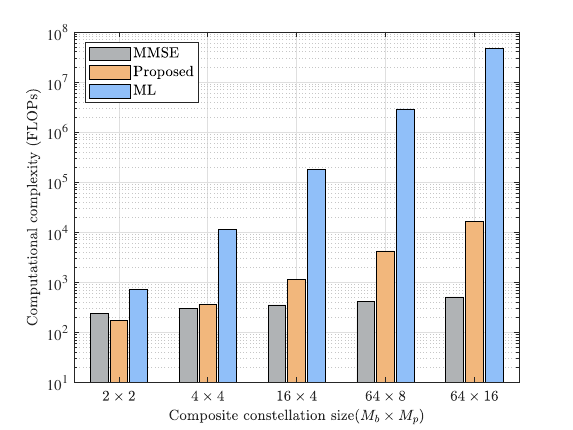}}
\caption{Complexity of different detectors under several composite constellation sizes ($M_b \times M_p$).}
\label{comlex}
\end{figure}

Fig.~\ref{comlex} illustrates the computational complexity of different detectors in terms of floating-point operations (FLOPs) as a function of the composite constellation size $M_b \times M_p$. Each group of bars corresponds to one constellation configuration, while the three bars within a group represent the MMSE detector, the proposed VAMP-based detector, and the ML detector, respectively. As expected, the complexity of all schemes increases with the modulation order, but at very different rates. The ML detector exhibits an exponential growth and quickly becomes prohibitive when the composite constellation size exceeds, e.g., $16\times 4$, being several orders of magnitude more complex than the other schemes for $64\times 8$. In contrast, the LMMSE and VAMP detectors scale only polynomially with $M_b$ and $M_p$, leading to much slower growth in FLOPs. The proposed VAMP-based detector incurs a moderate complexity overhead compared with MMSE but remains orders of magnitude less complex than ML across all considered configurations, thereby striking a favorable balance between performance and computational cost for PASM systems.

\begin{figure}[htbp]
\centerline{\includegraphics[width=0.7\linewidth]{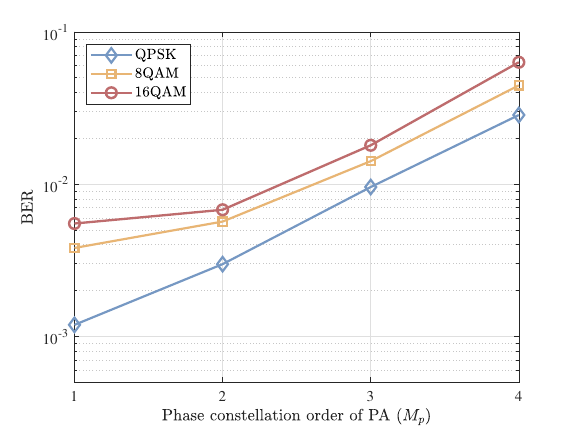}}
\caption{BER performance of the PASM system of phase‐shift constellation size for different baseband modulation orders.}
\label{phase}
\end{figure}

 Fig.~\ref{phase} illustrates the impact of the phase shift constellation size ($M_p$) on the BER performance of the PASM system for different baseband modulation schemes. For a fixed modulation order, the BER monotonically increases with $M_p$. This is because enlarging $M_p$ expands the composite constellation per waveguide and reduces the minimum Euclidean distance between constellation points, making the detector more sensitive to noise and residual interference. In addition, for any given $M_p$, higher-order baseband modulation (e.g., 16QAM) consistently exhibits worse BER than lower-order modulation (e.g., QPSK), owing to its denser symbol mapping. These results clearly reveal a fundamental tradeoff: increasing $M_p$ improves spectral efficiency by conveying more phase information but inevitably degrades error performance, so $M_p$ should be chosen according to the target reliability requirement.

\section{Conclusion}
In this paper, we proposed a PASM architecture that leverages dielectric waveguides and PAs to realize phase-shifter-free spatial multiplexing with a small number of RF chains. We established a physically grounded channel model and revealed a waveguide-level composite constellation that induces a block-structured prior on the transmit vector. To exploit this structure, we designed a VAMP-based detector in which an LMMSE module is coupled with a waveguide-structured denoiser that jointly processes all PAs on each waveguide. We also derived an analytical upper bound on the BER under ML detection and verified its tightness via simulation. Numerical results showed that PASM yields substantial SNR gains over conventional PSSM at the same spectral efficiency, while the proposed VAMP detector significantly outperforms linear baselines with polynomial-time complexity that is orders of magnitude lower than that of ML detection.

\ifCLASSOPTIONcaptionsoff
  \newpage
\fi





\bibliographystyle{IEEEtran}
\bibliography{IEEEabrv,Bibliography}

\vfill


\end{document}